\documentclass[letter]{aa}  

\usepackage{color}
\usepackage{graphicx}
\usepackage{natbib}
\usepackage[para,online,flushleft]{threeparttable}

\usepackage{txfonts}
\begin{document}

   \title{First detection of CF$^+$ towards a high-mass protostar}


   \author{S. Fechtenbaum
          \inst{1,2}
          \and
          S. Bontemps \inst{1,2}
          \and
          N. Schneider \inst{1,2}
          \and
          T. Csengeri \inst{3} 
            \and
            \\
          A. Duarte-Cabral \inst{4} 
            \and
          F. Herpin \inst{1,2} 
            \and
          B. Lefloch \inst{5}       
          }

   \institute{Univ. Bordeaux, LAB, UMR 5804, F-33270, Floirac, France\\
              \email{Sarah.Fechtenbaum@obs.u-bordeaux1.fr}
         \and
            CNRS, LAB, UMR 5804, F-33270 Floirac, France
         \and
             Max-Planck-Institut f\"ur Radioastronomie, Auf dem H\"ugel 69, 53121 Bonn, Germany
          \and
          School of Physics and Astronomy, University of Exeter, Stocker Road, Exeter EX4 4QL, UK
          \and
          Laboratoire AIM Paris-Saclay, CEA/IRFU - CNRS/INSU - Universit\'{e} Paris Diderot, CEA-Saclay, Gif-sur-Yvette Cedex, France
             }

\date{Received ---, --- ; accepted --- ---, ----}

\mail{sarah.fechtenbaum@obs.u-bordeaux1.fr}  

 
  \abstract
   {}
   {We report the first detection of the J = 1 - 0 (102.6 GHz) rotational lines of CF$^+$
     (fluoromethylidynium ion) towards CygX-N63, a young and massive protostar of the
     Cygnus X region.}
   {This detection occurred as part of an unbiased spectral survey of this object in the $0.8-3$ mm range, performed with the IRAM 30m telescope. The data were analyzed using a local thermodynamical equilibrium model (LTE model) and a population diagram in order to derive the column density.
   }
   {
     The line velocity (--4 km s$^{-1}$) and line width (1.6 km s$^{-1}$) indicate an origin from the collapsing envelope of the protostar.  We obtain a CF$^+$ column density of 4$\times$10$^{11}$cm$^{-2}$. The CF$^+$ ion is thought to be a good tracer  for C$^+$ and assuming a ratio of 10$^{-6}$ for CF$^+$/C$^+$, we derive a total number of C$^+$  of $1.2\times$10$^{53}$ within the beam. There is no evidence of carbon ionization caused by an exterior source of UV photons suggesting that the protostar itself is the source of ionization. Ionization from the protostellar photosphere is not efficient enough. In contrast, X-ray ionization from the accretion shock(s) and UV ionization from outflow shocks could provide a large enough ionizing power to explain our CF$^+$ detection. 
     }
   {Surprisingly, CF$^+$ has been detected towards a cold, massive protostar with no sign of an external photon dissociation region (PDR), which means that the only possibility is the existence of a significant inner source of C$^+$. This is an important result that opens interesting perspectives to study the early development of ionized regions and to approach the issue of the evolution of the inner regions of collapsing envelopes of massive protostars. The existence of high energy radiations early in the evolution of massive protostars also has important implications for chemical evolution of dense collapsing gas and could trigger peculiar chemistry and early formation of a hot core. }


   \maketitle
%

\section{Introduction}

The formation of massive stars is still not well understood. Important questions are related to accretion rates required to form massive stars in the monolithic collapse scenario, and to the evolution of the inner regions of the collapsing cores where the disks form and stellar feedback occurs \citep[e.g.][]{yorke2002,
  duarte-cabral2013}.  For high accretion rates (10$^{-4}$ to 10$^{-3}$~M$_\odot /$yr), the protostars have large radii (10 to 100 R$_\odot$) before they contract to reach the main sequence while still accreting \citep[e.g.][]{hosokawa2009}. It is roughly when the protostars reach the main sequence that they start to develop ultra-compact (UC H{\small II}) regions which may then regulate and possibly stop further accretion, limiting the final mass of the stars. Massive protostars also form hot cores with large fractions of collapsing gas radiatively warmed over 100 K. They release into the gas phase molecular species previously frozen on grains and drive a warm gas chemistry. It is not clear whether this hot core phase occurs before or at the same time as the development of an UC$\,$H{\small II} region. By opening large inner cavities, outflows and UC$\,$H{\small II} regions may help to heat up a large fraction of collapsing envelopes, leading to well-developed hot cores. 

CygX-N63 has been determined by \citet{motte2007} to be one of the brightest and most compact 1.2mm cores in the Cygnus X region \citep{schneider2006} at a distance of 1.4$\,$kpc \citep{Rygl2012}. It is very likely a massive protostar in its earliest phase of formation \citep{bontemps2010}. With an envelope mass of 44 M$_\odot$ within 2500 AU, and a luminosity of 340 L$_\odot$, it is the most massive and youngest protostar detected in Cygnus X, and is recognized as a Class 0 massive protostar driving a powerful CO outflow \citep{duarte-cabral2013}. It does not yet excite an UC$\,$H{\small II} region and does not contain a developed hot core and could thus be a rare example of a massive protostar in its pre-UC$\,$H{\small II} region and pre-hot core phase. This exceptional object is one of the best known examples of a dense core in monolithic collapse with no sign of sub-fragmentation from 0.1 pc down to 500 AU, although it could still form a close binary.

Molecular lines are important probes of the dense gas from which stars form and hence provide important insight for understanding the earliest phases of star formation. One of the major cooling lines of the warm interstellar medium is the far-infrared fine structure line of ionized carbon (C$^+$) at 158 $\mu$m.  It was recognized (\citealp{neufeld2005}) that in the molecular gas phase of the interstellar medium, HF is the main fluor reservoir and that CF$^+$ could be a good indirect tracer of C$^+$, which is not reachable from ground-based facilities. The first (and only) detections of mm-transitions of CF$^+$ towards the Orion Bar \citep{neufeld2006} and the Horsehead nebula (Guzman et al. 2012) confirmed these predictions and indicate that CF$^+$ can trace C$^+$ in photon dominated regions (PDRs).

Here we present the first detection of CF$^+$ towards a protostar obtained within a 0.8 to 3$\,$mm unbiased spectral survey employing the IRAM 30m towards CygX-N63. In Sect.~2 the observations are detailed while results are provided in Sect.~3. Sect.~4 discusses the origin of CF$^+$ detection and its
implications.


\begin{table*}
\begin{center}
\begin{threeparttable}
\small
\caption{\small{Observation parameters, Gaussian fits results and LTE model results. \label{observation}}}
\begin{tabular}{llllllllllll}
\\
 \hline
  \hline
Line & Frequency & E$_{\rm up}$/k & A$_{\rm ij}$ & F$_{\rm eff}$ & B$_{\rm eff}$ & Beam  & V$_{\rm LSR}$ & FWHM & T$_{\rm peak}$ & Noise
 \\
 & GHz &K & s$^{-1}$ & & & arcsec & km s$^{-1}$ & km s$^{-1}$ & mK & mK
 \\
 \hline
CF$^+$(1-0) & 102.587533 & 4.9235  & 4.82$\times$10$^{-6}$ & 0.94  & 0.79 & 25.5 &  -4.1 $\pm$ 0.5\tnote{a} & 1.6 $\pm$ 0.5\tnote{a}  & 10.5\tnote{a} & 2.3   \\
CF$^+$(2-1) & 205.170520 & 14.7703 & 4.62$\times$10$^{-5}$ & 0.94  & 0.64 & 11.4 &  -4.1\tnote{b}  & 1.6\tnote{b} & <62\tnote{c}  &  14.7 \\
 \hline
\end{tabular}
\tiny
\begin{tablenotes}

\item[a] Gaussian fit results \\
\item[b] Fixed equal to the J = 1 - 0 line parameters \\
\item[c] Upper limit. LTE modeling with a 25'' source size (beam size) and an excitation temperature of 10 K gives T$_{\rm peak} = 24\,$mK.\\
\end{tablenotes}

\end{threeparttable}
\end{center}
\end{table*}

\section{Observations and data reduction}

We carried out an unbiased spectral survey with the IRAM 30 m telescope on Pico Veleta, Spain, towards the source CygX-N63, located at (RA, Dec)$_{\rm{J2000}}$ = (20$^h$40$^m$05.2$^s$, 41$^{\circ}$ 32$'$ 12.0$''$). CygX-N63 is an isolated single object and can therefore be observed without confusion within the beam of the IRAM 30m (up to 30$^{\prime\prime}$ at 3mm). The observations were obtained between September 2012 and January 2014 using the wobbler switching mode to assure good baselines, and covering so far the whole 1, 2, and 3 mm bands ($80-270\,$GHz).  The EMIR receiver was connected to the FTS backend at 195 kHz with settings, separated by 3.89 GHz to cover the bands with a redundancy of 2. In order to distinguish ghosts from bright lines in the rejected sideband, we observed each frequency as pairs of settings shifted by 20 MHz. The data were reduced with the GILDAS package\footnote{http:/www.iram.fr/IRAMFR/GILDAS/}. After a zeroth-order baseline , all spectra were averaged and converted into main-beam temperature (T$_{\rm mb}$) using the forward and main-beam efficiencies (F$_{\rm eff}$ and B$_{\rm eff}$) listed in Table ~\ref{observation} leading to typical rms noise levels of 2.9, 4.2, and 8.6 mK for the 3, 2, and 1 mm band, respectively. 

 \begin{figure}[h!]
  \begin{center}
    \includegraphics[width=0.75\linewidth]{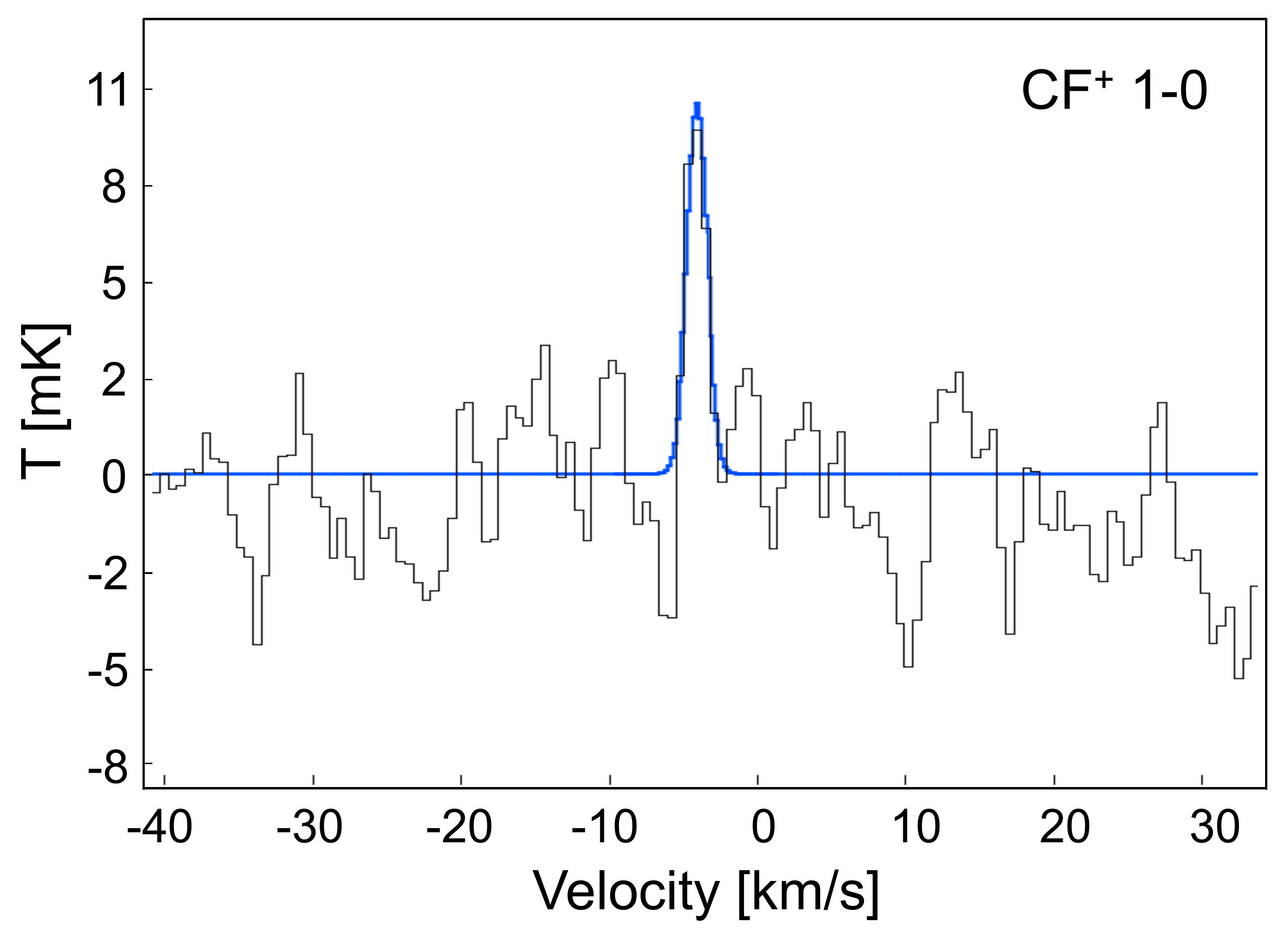}
       \includegraphics[width=0.75\linewidth]{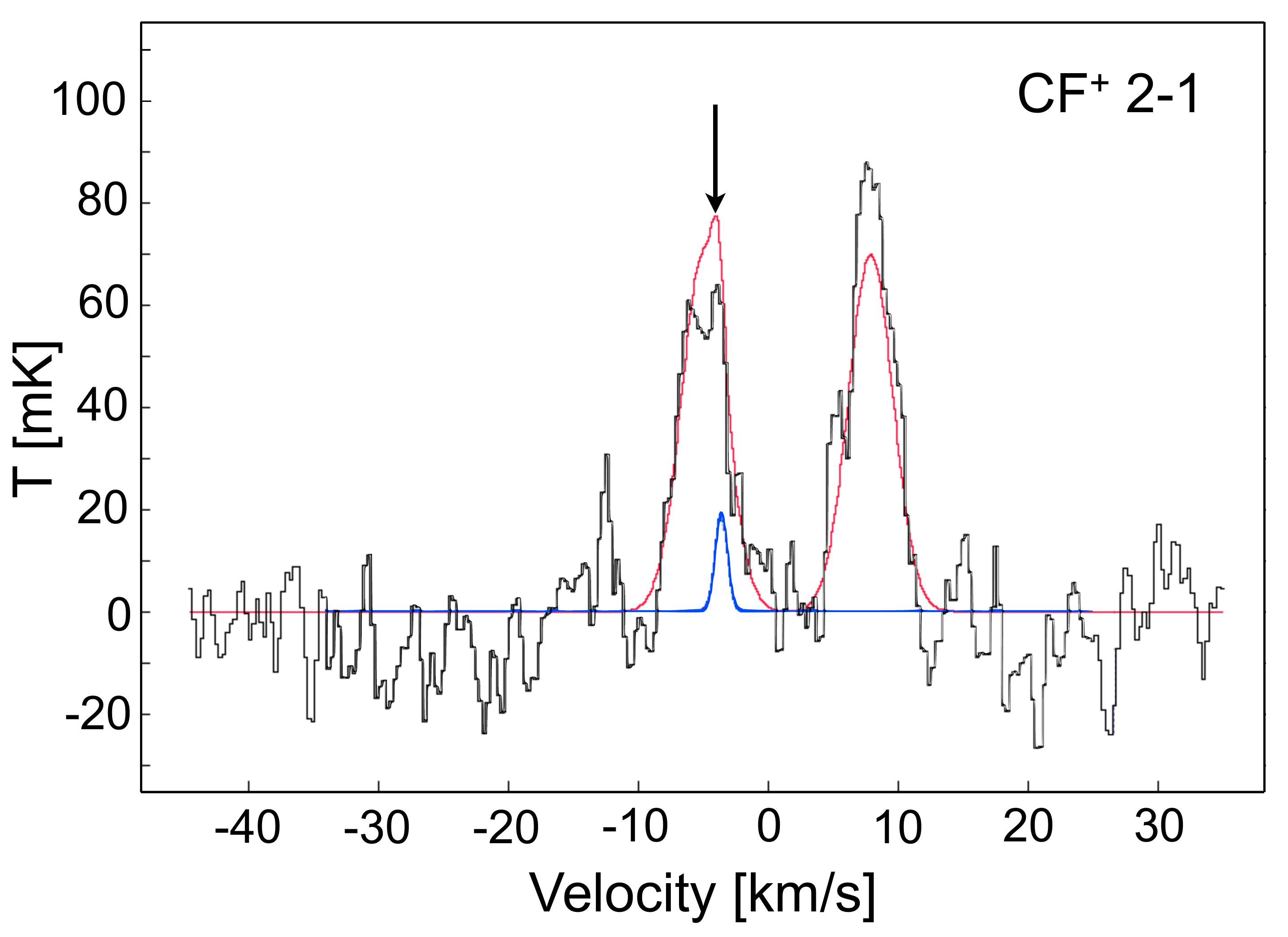}
       \caption{\small{Spectra of the CF$^+$ J = 1$\to$0 and 2$\to$1
           lines on a main beam brightness temperature scale. The
           blue curves are the modeled spectra of the CF$^+$ lines and the red curve is the model for CF$^+$ and CH$_3$CHO together. } }
    \label{spectra}
  \end{center}
\end{figure}

\section{Analysis and results}

\subsection{Line identification and detection of CF$^+$ (1-0)}

The analysis was performed using CASSIS\footnote{http://cassis.irap.omp.eu/ } and the CDMS\footnote{http://www.astro.uni-koeln.de/cgi-bin/cdmssearch} catalogue. We here report a clear detection (6.2$\,\sigma$) of the J = 1$\to$0 line of CF$^+$ at 102.59 GHz at $v_{\rm LSR} = -4\,$km/s. The only other line of CF$^+$ (J = 2$\to$1 at 205.17 GHz) in our spectral coverage cannot be firmly detected because of blending with a line of CH$_3$CHO (11$_{1\ 11\ 0}$ - 10$_{1\ 10\ 0}$) (see Fig.~\ref{spectra}). 

\subsection{Total column density and excitation temperature}

Using an excitation temperature of 10 K as in \citet{neufeld2006} and \citet{guzman2012}, assuming local thermodynamical equilibrium (LTE) and optically thin lines, we derive a total CF$^+$ column density of  $4\times$10$^{11}$cm$^{-2}$ (see Appendix A for details). Taking the (2$\to$1) line tentative detection as an upper limit, we obtain a maximum excitation temperature of $\sim 20\,$K.

The line velocity of --4 km s$^{-1}$ is exactly the rest velocity of the protostar \citep{bontemps2010}, which suggests an origin from the protostar. The study of line profiles suggests a correlation between the line width and the spatial origin of molecules  (Fechtenbaum et al. in prep.). The observed full width at half maximum (FWHM) of 1.6 km s$^{-1}$ points to an origin from the envelope.

\section{Discussion and conclusion}

\subsection{First detection of CF$^+$ towards a massive protostar} \label{theory}

The CF$^+$ ion is stable and is the second fluorine reservoir after HF in PDRs. Its formation route is simple \citep{neufeld2005}

F + H$_2$ $\rightarrow $ HF + H           \hspace{0.6cm} and     \hspace{0.6cm}     HF + C$^+$ $\rightarrow $ CF$^+$ + H,

\noindent
where both HF and C$^+$ are abundant, i.e. at the interface H{\small I}$\,-\,$H$_2$ and more generally where ionization is strong enough to produce C$^+$. CF$^+$ should be an excellent proxy for C$^+$ in partially ionized regions.  CF$^+$ has so far been detected only towards two bright PDRs: the Orion Bar and the Horsehead nebula. The Orion Bar is an intense PDR partly seen edge-on leading to a large column density of CF$^+$ \citep{neufeld2006}. The Horsehead nebulae is a weaker PDR, but the interface is seen almost perfectly edge-on with a significant limb-brightening increasing the observed column density of CF$^+$ \citep{guzman2012}. 

The present detection is the first detection towards an object which is most likely not a PDR (see Sect.~\ref{pdr}). It is an interesting finding because it then points to the existence of a source of C$^+$ inside the CygX-N63 core. A column density of $4\times10^{11}\,$cm$^{-2}$ in a 25$^{\prime\prime}$ FWHM beam corresponds to a total number of CF$^+$ ions of $1.2\times10^{47}$. With the abundance ratio CF$^+$/ C$^+$ of the order of $10^{-6}$ at high density and high UV field \citep[see Fig. 4 in][]{neufeld2006}, it converts to a total number of C$^+$ of $1.2\times10^{53}$. We discuss below the different possibilities to explain this amount of C$^+$ indirectly traced by CF$^+$.

   \begin{figure}
   \hspace{-0cm}\includegraphics[angle=0,width=8.0cm]{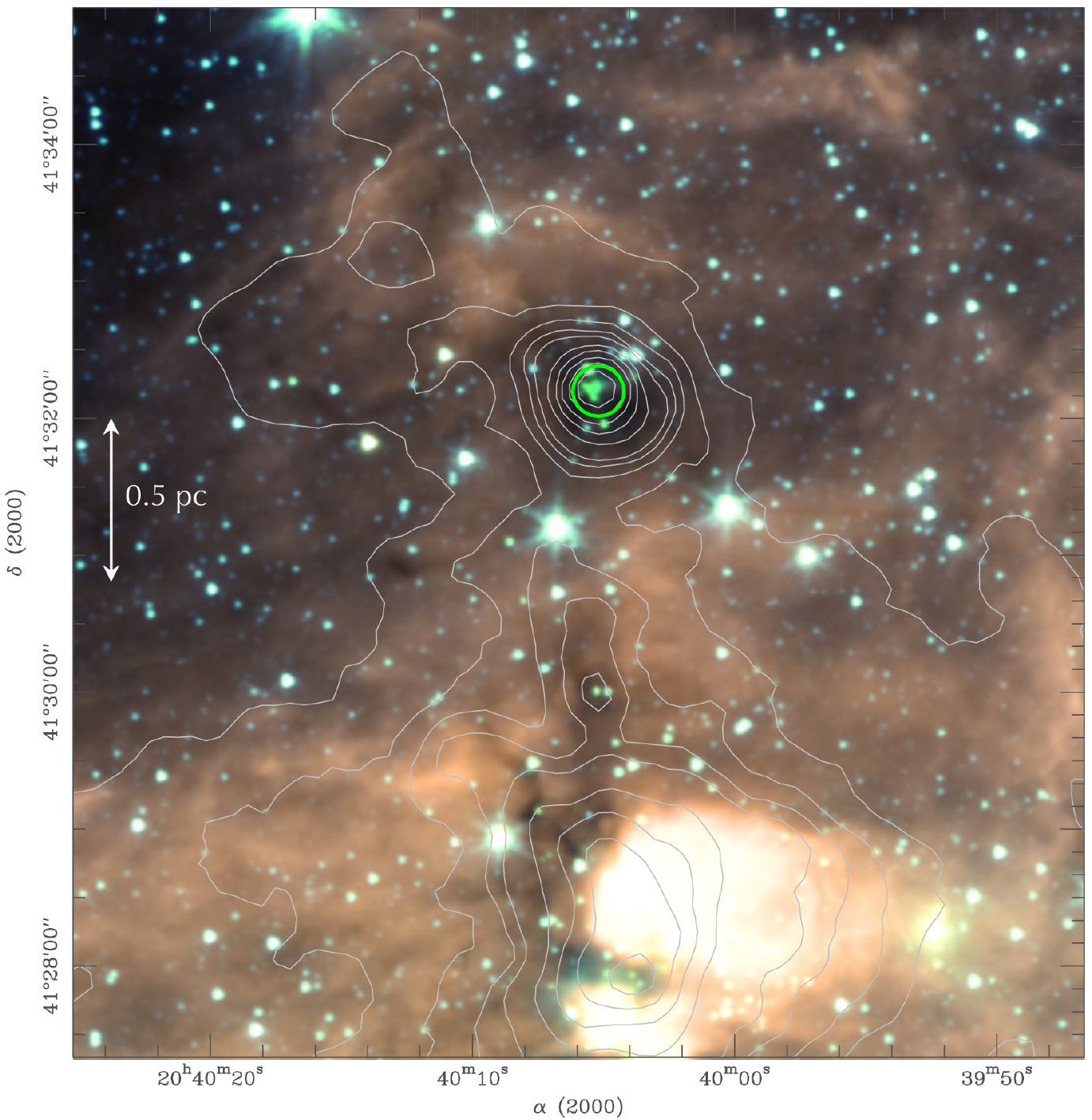}
   \vspace{-0cm}
      \caption{ IRAM 30m beam at 102 GHz (green circle) overlaid on a RGB (8, 4.5, and 3.6$\,\mu$m) Spitzer IRAC image towards CygX-N63. Contours show the 250$\,\mu$m dust emission from Herschel \citep{henneman2012}. No prominent PAH emission (red, 8$\,\mu$m) is detected towards CygX-N63.
   }
         \label{IRAC}
   \end{figure}

\subsection{A PDR front detected at the distance of Cygnus X?}
 \label{pdr}

Since CygX-N63 is situated in the rich Cygnus X complex hosting a large number of OB stars, the detected CF$^+$ could potentially originate from the PDR at the surface of the clump hosting CygX-N63, excited by the most nearby massive stars. However, CygX-N63 is located in a quiet part of Cygnus X where no nearby O stars can be detected (the projected distance to the central regions of the Cygnus OB2 cluster is $\sim$20-30 pc). In comparison, the Orion Bar PDR is situated at less than 0.5 pc from the O stars of the Orion nebula cluster. To detect such a weak CF$^+$ line in Cygnus X, a PDR has to be bright and should therefore exhibit strong PAH emission in the mid-IR. In Fig.~\ref{IRAC} showing a three-color Spitzer image, one sees that the CygX-N63 core actually coincides with one of the darkest parts of the region with no sign of PAH red (8$\,\mu$m) emission in the observed beam. Only some green (4.5$\,\mu$m) weak extended features are seen precisely coinciding with CygX-N63. It does not trace a PDR front and is most probably due to outflow shock \footnote{There are no PAH features at 4.5$\,\mu$m. Such strong excesses at 4.5$\,\mu$m are usually referred to as EGOs (extended green objects) and are due to H$_2$ line emission in outflow shocks \citep[e.g.][]{cyganowski2008}.} from the strong CO outflow driven by CygX-N63 (see Sect.~\ref{outflow}). There is therefore no hint of the existence of a bright PDR at the surface of the clump which would explain the observed strong source of C$^+$.

\subsection{An inner CII region towards CygX-N63?} 
\label{cii}

CygX-N63 is an OB star in formation. It is or will soon become a strong ionizing source and a strong source of C$^+$. Interestingly enough, however, CygX-N63 is not yet an UC$\,$H{\small II} region. It has not been detected in free-free emission with the VLA. CygX-N63 only coincides with a $2.3\,\sigma$ peak of 0.14 mJy at $8.4\,$GHz (see Appendix~\ref{vla-obs}). Since carbon has a lower ionization energy (11.3 ev) than hydrogen (13.6 ev), a C$^+$ region can exist before the development of an H{\small II} region. 

With a luminosity of 340 L$_\odot$, CygX-N63 can host a B5$\,-\,$B6 main sequence star of photospheric temperature 15000 K with a carbon ionizing flux (between 11.3 and 13.6 ev) of $2.63\times10^{45}\,$s$^{-1}$ \citep{Diaz-Miller1998}. Assuming a r$^{-2}$ law and having 44.3 M$_\odot$ inside a FWHM of 2500 AU \citep{duarte-cabral2013}, we get an H$_2$ density of $1.95\times10^{10}\times({\rm r}/100\,{\rm AU})^{-2}\,$cm$^{-3}$. Using the recombination rate of the UMIST database (\citealp{McElroy2013}), assuming all carbon in atomic form and that ionization is dominated by carbon, we can estimate the thickness of the CII region (see Appendix~\ref{r-rate} for details). For this estimate we also neglect the effect of dust extinction on the ionizing radiation. Even with these strong assumptions the above carbon ionizing flux can ionize a thin layer of only 0.1 AU for an internal radius of 100 AU (Appendix~\ref{r-rate}). Such a thin layer would only contain $2.6\times10^{50}$ ionized carbons, i.e. $\sim 500$ times less than observed. To increase the efficiency of the ionization, the ionizing radiation needs to reach lower density layers where the recombination is less efficient so that the relatively weak ionizing radiation is able to keep a large amount of carbon ionized. The typical required H$_2$ density is $4\times10^{7}\,$cm$^{-3}$ (see Appendix~\ref{r-rate}). Such a density is expected at large radii, of the order of 2000 AU, which cannot be easily reached by inner UV radiation because of the expected large dust extinction from the inner envelope. Since CygX-N63 drives an outflow \citep{duarte-cabral2013}, its envelope has bipolar cavities, and should have a disk-like structure in the central regions that reduces extinction in the equatorial directions. It is difficult, however, to expect a reduced enough extinction for ionizing radiation to reach 2000 AU, which is close to the physical size of the envelope (see \citealp{bontemps2010}).

\subsection{X-ray ionization from an accretion shock?}
\label{accretion}

CygX-N63 is actually a young protostar (Class 0) with a luminosity dominated by accretion. An accretion shock creates a hot plasma which radiates half of the accretion power outward. This plasma has a typical temperature of 10$^6\,$K and mostly radiates in extreme UV, but also in X-rays. X-rays ionize gas more efficiently than does photospheric radiation (e.g. \citealp{Montmerle2001}), and are also less affected by extinction than UVs \citep{ryter96} and could thus ionize carbon deeper in the envelope decreasing the extinction problem discussed in the previous section. A large fraction of X-ray ionization originates from locally emitted UV radiation (possibly down to carbon ionizing photons) as a result of local deposition of energy from photoelectric electrons generated by X-rays \citep{Krolik&Kallman1983,Glassgold2000}. This has the interesting effect that X-rays, especially high-energy X-rays, can propagate deep in the envelope, being less affected by extinction than UVs, and deposit locally their energy as UV ionizing radiation. The total luminosity of CygX-N63 could represent up to $7.2\times10^{46}\,$s$^{-1}$ ionizing photons for carbon, assuming that all the radiated energy is dedicated to C ionization, which is obviously the most optimistic view. Using the same assumptions as in Sect.~\ref{cii}, we find that the carbon ionized layer for an inner cavity of 100 AU would then be 2.6 AU and would contain $6.8\times10^{51}$ ionized carbons, i.e. $\sim 20$ times less than observed. This is still a bit low, but the discrepancy is reduced compared to Sect.~\ref{cii}. The local density required to keep $1.2\times10^{53}$ carbons ionized would be $1.1\times10^{9}\,$cm$^{-3}$, i.e. $\sim 20\,$ times lower than the density at 100 AU in the pure spherical case. If the reachable density could be a bit reduced, for instance at the surface of the outflow cavities, as in the irradiated outflow wall scenario of \citet{bruderer2009}, the agreement could perhaps be reached, providing that a significant fraction of the available energy can be transfered to carbon ionization as discussed above. The estimate of the possible X-ray luminosity which could increase by orders of magnitude during accretion bursts (see \citealp{Montmerle2001} and references therein) and the overall efficiency of X-ray ionization at $R\sim100\,$AU accounting for X-ray absorption in such a complex geometry are difficult to assess precisely. We can only conclude that the ionizing power expected from an accretion shock may reach the right order of magnitude to explain our CF$^+$ detection.

\subsection{Outflow shocks?} 
\label{outflow}

The last possible source of ionization is outflow shocks whose energy is extracted from the inner regions of the collapsing envelope. With a CO velocity of 40~km/s and an outflow force of $2.9\times10^{-3}\,$M$_\odot \,$km/s$\,$yr$^{-1}$ \citep{duarte-cabral2013}, we get an outflow luminosity of 9.6$\,$L$_\odot$ which could represent up to $2\times10^{45}\,$s$^{-1}$ carbon ionizing photons. Young jets are expected to radiate their energy through shocks. Three knots of 4.5 $\mu$m emission (shocked H$_2$) are indeed located in projection at only $\sim 3.5^{\prime\prime}$ from CygX-N63 (Fig.~\ref{IRAC}). These shocks should radiate in UV \citep{vanKempen2009}. For an average projection angle of 57$^\circ$, the shocks would be then located at a radius of $9000\,$AU from the protostar where the expected $n_{\rm H}$ density in the envelope is $4.8\times10^{6}\,$cm$^{-3}$ (see Sect.~\ref{cii}).  At this density an ionizing flux of only $1.6\times10^{44}\,$s$^{-1}$ is required to keep ionized $1.2\times10^{53}$ carbons. This is less than the available ionizing power of $2\times10^{45}\,$s$^{-1}$.  An attenuation of the ionizing UVs in the outflow cavities is expected, but it is reduced to only 0.24~mag for the expected density\footnote{For pressure equilibrium \citep{bruderer2009} and infall and outflow velocities of 1 and 40 km/s respectively, the cavity density should be (1/40)$^2$ of $4.8\times10^{6}\,$cm$^{-3}$, i.e. $3.0\times10^{3}\,$cm$^{-3}$} of $3.0\times10^{3}\,$cm$^{-3}$ in the cavities and for a distance\footnote{For an outflow opening angle of 60$^\circ$, the distance to the outflow walls is typically of 4500~AU. For $n_{\rm H} = 3\times10^{3}\,$cm$^{-3}$, the visual extinction $A_{\rm V}$ is then of the order of 0.1~mag and the UV extinction of 0.24~mag using the $A_{\rm UV}/A_{\rm V}$ ratio of 2.4 used in \citet{bruderer2009}.} of 4500 AU. The shocks in the outflow may thus provide the required amount of ionizing photons to explain our CF$^+$ detection.

\subsection{A CII region without HII region} 
\label{cii-hii}

In the two possible scenarios discussed above (Sects.~\ref{accretion} and \ref{outflow}) the produced UV radiation should ionize first hydrogen and create an H{\small II} region which has, however, not been detected in free-free emission (Fig.~\ref{VLA}). For densities and sizes involved in the accretion shock case the free-free emission has to be optically thick. We then estimate that the 5$\,\sigma$ upper limit in free-free of 0.3 mJy would correspond to an emitting region of $\sim 20\,$AU radius (see Appendix~\ref{ff}). The C ionized region is expected to be more extended than the HII region and given the very uncertain geometry and density distribution, 20 AU could be considered to be roughly compatible with the $100\,$AU scale discussed above.
In the outflow shock scenario, for a dissociative outflow shock with a jet velocity of 200 km/s and $n_{\rm H}=3\times 10^{-3}\,$cm$^{-3}$, the free-free emission is optically thin with $\tau_{\rm ff}=5\times10^{-4}$.  For an ionizing flux of $2\times10^{44}\,$s$^{-1}$, using a recombination rate of $3\times10^{-13}n_{\rm  e-}\,$s$^{-1}$ \citep{spitzer1978} and for full ionization ($n_{\rm  e-}=3\times 10^{-3}\,$cm$^{-3}$), one gets EM$_{\rm V}=6.6\times10^{56}\,$cm$^{-3}$ leading to $0.94\,$mJy at 8.4$\,$GHz (see Appendix~\ref{ff}). This is only three times larger than the 0.3 mJy upper limit with the VLA which could be seen as a still acceptable discrepancy given the large uncertainties in the above estimates. We can therefore conclude that in both the accretion and outflow shocks scenarios a CII region can be detected in CF$^+$, while no free-free emission has been detected in the accompanying HII region.

\subsection{A new probe of the early evolution of massive protostars} 

In conclusion, the first detection of CF$^+$  towards a massive protostar may provide us with a new tool for investigating the ionizing power of accretion and outflow shocks. In contrast to C$^+$ in the far-IR, CF$^+$ in the millimeter range can be observed at high spatial resolution with interferometer. It could offer the opportunity to image the development of C$^+$ regions even in the case of non-detection of any $\,$H{\small II} region. The indication of a possible strong X-ray and/or UV ionization also has important implications for the early chemical evolution in collapsing envelopes in the context of the formation of a hot core.

\begin{acknowledgements}
We thank V. Wakelam for fruitful discussions and N. Lagarde for reducing VLA data. This work is supported by the IdEx Bordeaux funding and by the project STARFICH, funded by the French National Research Agency (ANR). ADC acknowledges funding from the FP7 ERC starting grant project LOCALSTAR.
\end{acknowledgements}


\bibliographystyle{unsrt}

\appendix

\section{CASSIS LTE modeling}

CASSIS provides a Jython script which minimizes the $\chi^2$ to find the best parameters to adjust the observations. With a spectrum i of N points, the $\chi^2$ is derived as
$$
\chi^2_i = \sum^{\rm{N}}_{\rm{j=1}}\frac{(\rm{I}_{\rm{obs,j}} - \rm{I}_{\rm{model,j}})^2}{\rm{rms}^2_i+ cal^2_i(\rm{I}_{\rm{obs,j}}-\rm{I}_{\rm{cont,j})^2}},
$$
\noindent where I$_{\rm{obs}}$ and I$_{\rm{model}}$ are respectively the observed and the modeled intensities, I$_{\rm{cont}}$ is the continuum intensity, cal$_i$ the calibration uncertainty for the spectrum i and N is the total number of points of the spectrum.
The reduced $\chi^2$ is the calculated as
$$
\chi^2_{\rm{red}}= \frac{1}{\rm{N_{spectra}}}\sum^{\rm{N_{spectra}}}_{\rm{i=1}}\frac{\chi^2_i}{\frac{\rm{N}}{\rm{N_{ind}}}-\rm{dof}},
$$
\noindent with N$_{\rm{spectra}}$ the number of spectra, N$_{\rm{ind}}$ the number of independent points, and dof the degree of freedom.

The adjustable parameters are the column density, temperature, FWHM, size of the source, and v$_{\rm{LSR}}$. In the case of CF$^+$, the column density varies between 10$^{10}$ and 10$^{14}$ cm$^{-2}$ with 500 steps; the other parameters were fixed. The temperature was set to 10 K, as in \citet{neufeld2006} and \citet{guzman2012}. The FWHM was measured on the spectrum as 1.6 km/s. No source size was assumed, so we kept the 30 m beam size of 25''. The velocity of the source is around -4 km/s. CASSIS also takes into account the beam size variation as a function of the frequency. The obtained $\chi^2_{\rm{red}}$ is 3.22 and the column density is (4 $\pm$ 1) $\times$ 10$^{11}$ cm$^{-2}$.

The V$_{\rm{LSR}}$ and FWHM derived for the J = 1 - 0 line are then used to predict the intensity of the J = 2 - 1 line. We detected 202 CH$_3$CHO lines in the whole coverage and used a population diagram to derive a mean excitation temperature of 40 K and a column density of 3 $\times$ 10$^{13}$ cm$^{-2}$. These values were used to model the CH$_3$CHO lines as shown in the bottom panel of Fig. \ref{spectra}.

\section{VLA observations}
\label{vla-obs}

\begin{figure}
  \hspace{-0cm}\includegraphics[angle=0,width=8.5cm]{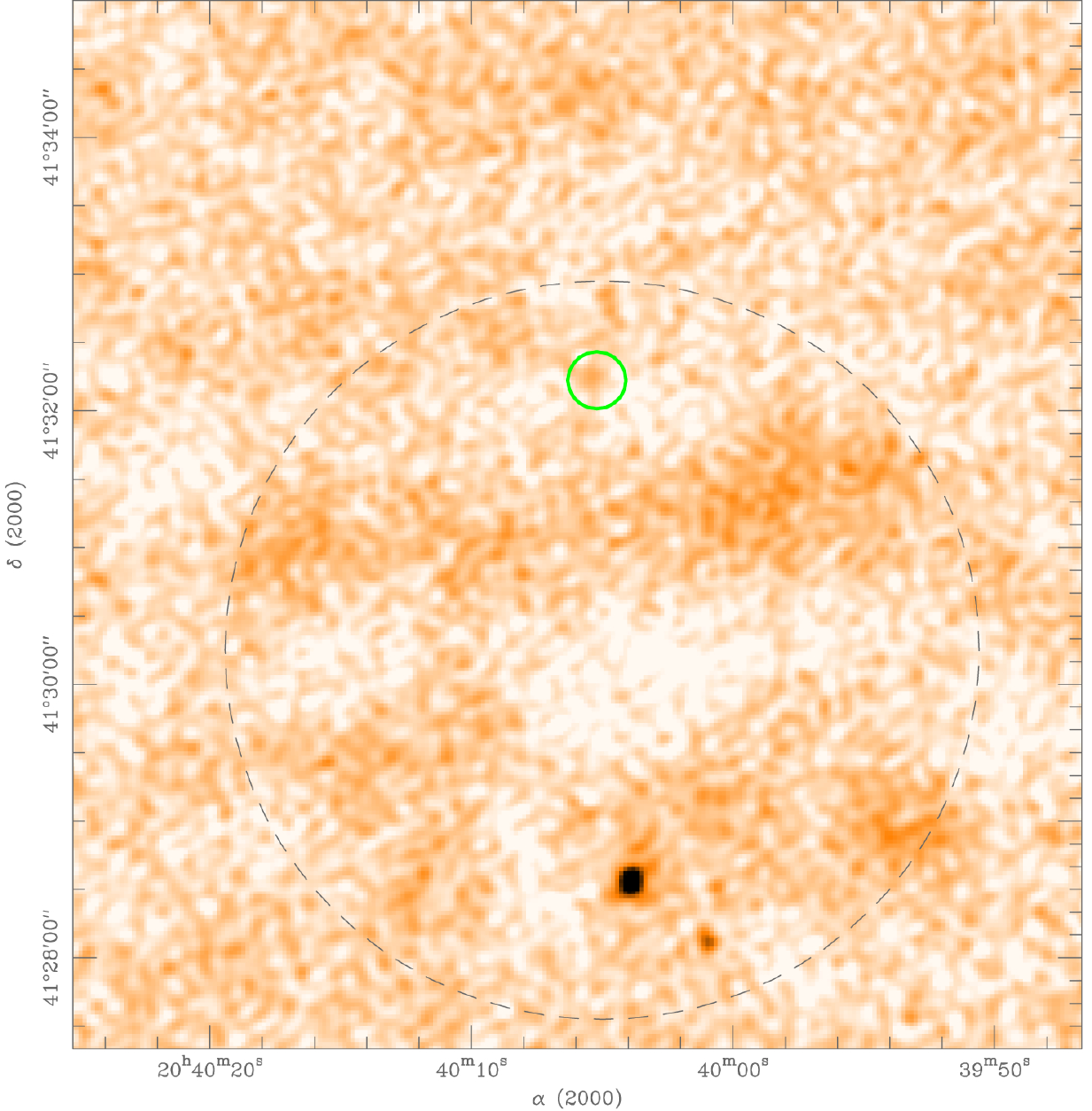}
   \vspace{-0cm}
     \caption{VLA image at 8.4 GHz of the same region than in Fig.~\ref{IRAC} with color scale from -0.1 to 0.7 mJy/beam. The dashed ellipse indicates the VLA primary beam and the green circle shows the IRAM 30m beam at 102 GHz. The rms in this image stays large, of the order of 0.06 mJy/beam, due to strong sidelobes originating from to DR22, a strong radio source (a compact HII region) outside the imaged field ($\sim$5 arcmin south). 
              }
         \label{VLA}
   \end{figure}

CygX-N63 was observed on 27 April 2003 with the 27 antennas of the VLA in D configuration at 8.4 GHz (band X) (project AB1073). The image shown in Fig.~\ref{VLA} corresponds to a total integration time of 25 mins spread over 4.5 hr (track sharing technique to improve UV coverage). It has been cleaned with the clean algorithm of AIPS. The resulting rms is equal to 60.2 $\mu$Jy/beam.

In the direction of CygX-N63, a weak peak of 0.14$\,$mJy, i.e. at a level of 2.3$\,\sigma$, is found. Below 3$\,\sigma$ it cannot be considered as a possible detection, especially because the side lobes (large stripes in the image of Fig.~\ref{VLA}) are roughly of the same intensity. 

\section{Carbon ionization equilibrium}
\label{r-rate}

At equilibrium in spherical geometry from an initial radius $r_{\rm i}$ to the radius of the C{\small II} region (where mostly all carbons are ionized) $r_{\rm CII}$ the equilibrium of carbon ionization can be expressed as 
$$
Q_{\rm CII} = \int_{r_{\rm i}}^{r_{\rm CII}} k_{\rm rec}  n_{\rm  c+}  n_{\rm  e-} 4 \pi r^2 {\rm d}r,
$$

\noindent
where $Q_{\rm CII}$ is the total rate of ionizing photons  (in s$^{-1}$), $k_{\rm rec}$ is the carbon recombination rate that we take as equal to 2.36$\,\times\,10^{-12}(T/300)^{-0.29}$exp($-17.7/T$) cm$^3\,$s$^{-1}$ (UMIST 2012 database \citealp{McElroy2013}), and $n_{\rm  c+}$ and $ n_{\rm  e-}$ are the densities of C$^+$ and of electrons. For a large range of gas temperature (10 to 4000~K), $k_{\rm rec}$ stays within a factor of less than 2 around the adopted value of $1.74\,\times\,10^{-12}\,$s$^{-1}$ (ranging from 1.08 to 2.80$\,\times\,10^{-12}\,$s$^{-1}$ with the maximum obtained at $\sim 60\,$K). 

If carbon ionization dominates and assuming C$^+$/H equal to the standard C/H ratio of $1.6\times10^{-4}$ (if most of the carbon is in atomic form), we have $n_{\rm e-} = n_{\rm C+} = 1.6\times10^{-4}\times n_{\rm H}$. Assuming a r$^{-2}$ density law and having 44.3 M$_\odot$ inside a FWHM of 2500 AU \citep{duarte-cabral2013}, we here get an H$_2$ density of $1.95\times10^{10}\times({\rm r}/100\,{\rm AU})^{-2}\,$cm$^{-3}$. For this simple model, we get the following expression of the extent of the ionized region in the inner envelope,
$$
r_{\rm CII} = \frac{r_{\rm i}}{1-\eta\, r_{\rm i}} \hspace{0.4cm}{\rm with} \hspace{0.4cm} \eta= \frac{Q_{\rm CII}}{2.85\times10^{50}\,{\rm s}^{-1}},
$$
\noindent 
and the radii expressed in AU. For the case of Sect.~\ref{cii} ($Q_{\rm CII} =2.63\times10^{45}\,$s$^{-1}$), one gets a very small value of $9.4\times10^{-6}$ for $\eta$. For $r_{\rm i} = 100\,$AU,  $r_{\rm CII}$ is then equal to 100.1~AU. Only the very first layers are ionized for values of $Q_{\rm CII}$ significantly lower than $\sim 10^{50}\,$s$^{-1}$.

The ionization equilibrium can also be expressed as 
$$
Q_{\rm CII} = \int_{r_{\rm i}}^{r_{\rm CII}} k_{\rm rec}  n_{\rm  e-} {\rm d} N_{\rm  c+}
$$
with $N_{\rm  c+}$ being the number of ionized carbons. It shows that $k_{\rm rec}  n_{\rm  e-}$ is the fraction of ionized carbons that recombines per second. For $n_{\rm e-} = 1.6\times10^{-4}\times n_{\rm H}$, we can then derive the typical density (assuming constant density) required to keep an amount of ionized carbon $N_{\rm  c+}$ ionized under an ionizing rate $Q_{\rm CII}$. Here is the expression of this density $n_{\rm H}\!\!^{\rm CII}$ for our adopted numerical values:
$$
n_{\rm H}\!\!\!^{\rm CII} = 3.6\times10^{15}  \frac{Q_{\rm CII}}{N_{\rm  c+}}\,{\rm cm}^{-3}.
$$
\noindent 
For the case of Sect.~\ref{cii} with $Q_{\rm CII} =2.63\times10^{45}\,$s$^{-1}$ and $N_{\rm  c+} = 1.2\times10^{53}$, one gets $n_{\rm H}\!\!^{\rm CII}=7.9\times10^{7} \,$cm$^{-3}$.

\section{Centimetric free-free emission}
\label{ff}

As discussed in Sect.~\ref{cii-hii}, any ionized gas is expected to emit free-free emission.
 
The centimetric flux of free-free emission in the optically thin case can be expressed as a function of  EM$_{\rm V}$, the emission measure (see \citealp{Andre87}; \citealp{Curiel87}),
$$
\bigg(\frac{S_{\nu}}{{\rm mJy}}\bigg) = 1.41\,\,\bigg(\frac{EM_{\rm V}}{10^{57} \,{\rm cm}^{-3}}\bigg)\,\bigg(\frac{T_{\rm e}}{10^4\, {\rm K}}\bigg)^{-0.35}\,\bigg(\frac{\nu}{8.4\, {\rm GHz}}\bigg)^{-0.1}\,\bigg(\frac{d}{1.4\, {\rm kpc}}\bigg)^{-2}
$$
with
$$
EM_{\rm V} \equiv \int \! \! \! \int \!\! \!  \int  n_{\rm e}^2 \,{\rm d}V.
$$

\noindent
The optical depth is expressed as a function of the linear emission measure EM$_{\rm L}$ ($\equiv \int  n_{\rm e}^2 \,{\rm d}l$),
$$
\tau_{\rm \nu} \sim 1.22\,\,\bigg(\frac{EM_{\rm L}}{10^{27} \,{\rm cm}^{-5}}\bigg)\,\bigg(\frac{T_{\rm e}}{10^4\, {\rm K}}\bigg)^{-1.35}\,\bigg(\frac{\nu}{8.4\, {\rm GHz}}\bigg)^{-2.1},
$$

\noindent
which is equal to $\sim 8.8 \times 10^{6}$ for $n_{\rm e-}=n_{\rm H}=2.2\times10^{9}\,$cm$^{-3}$, $l=100\,$AU (see Sect.~\ref{accretion}), $10^4\,$K (the opacity is even larger for lower temperatures) and at $8.4\,$GHz (frequency of the VLA observation from Appendix~\ref{vla-obs}).

In the optically thick case the emission depends on the area of emission expressed in physical surface $A_{\rm em}$ if distance scaling is included,
$$
\bigg(\frac{S_{\nu}}{{\rm mJy}}\bigg) = 2.60\,\,\,\frac{A_{\rm em}}{(100\,{\rm au})^2}\,\,\,\bigg(\frac{T_{\rm e}}{10^4\, {\rm K}}\bigg)\,\,\,\bigg(\frac{\nu}{8.4\, {\rm GHz}}\bigg)^{2}\,\bigg(\frac{d}{1.4\, {\rm kpc}}\bigg)^{-2},
$$
\noindent
which is equal to 0.3~mJy for $A_{\rm em} = \pi R_{\rm em}\!\!\!^2$ with $R_{\rm em} = 19.2\,$AU (Sect.~\ref{accretion}).

\citet{Curiel87,Curiel89} also expressed the expected flux and optical depth in the case of a dissociative shock as a function of pre-shock density $n_0$ and shock velocity $V_{\rm s}$ (for $V_{\rm s}\gtrsim60\,$km/s),
$$
\tau_{\rm \nu} \sim 5.22\times10^{-4}\,\bigg(\frac{n_0}{10^4 \,{\rm cm}^{-3}}\bigg)\,\bigg(\frac{V_{\rm s}}{100\, {\rm km.s}^{-1}}\bigg)^{1.68}\,\bigg(\frac{{\rm T}_{\rm e}}{10^4\, {\rm K}}\bigg)^{-0.55}\,\bigg(\frac{\nu}{8.4\, {\rm GHz}}\bigg)^{-2.1},
$$
\noindent
which is found equal to $5\times10^{-4}$ for $n_0=3\times10^{-3}\,$cm$^{-3}$, $V_{\rm s}=200\,$km/s (see Sect.~\ref{outflow}), $10^4\,$K and at $8.4\,$GHz.

\end{document}